\documentclass[twocolumn,floatfix,superscriptaddress,amsmath,showpacs,showkeys,aps,prl]{revtex4-1}

\usepackage{t1enc}
\usepackage[final]{graphicx}
\usepackage{epsfig}
\usepackage{bm}
\usepackage[normalem]{ulem}
\usepackage[usenames]{color}
\usepackage{times}
\usepackage{verbatim}

\begin{document}

\title{Nature of long range order in stripe forming systems with long range repulsive interactions}

\author{Alejandro Mendoza-Coto}
\affiliation{Departamento de F\'\i sica, Universidade Federal do Rio Grande do Sul, CP 15051, 91501-970, Porto Alegre, Brazil}
\author{Daniel A. Stariolo}
\email{daniel.stariolo@ufrgs.br}
\affiliation{Departamento de F\'{\i}sica,
Universidade Federal do Rio Grande do Sul and
National Institute of Science and Technology for Complex Systems\\
CP 15051, 91501-970 Porto Alegre, RS, Brazil}
\author{Lucas Nicolao}
\affiliation{Departamento de F\'{\i}sica,
Universidade Federal de Santa Catarina, 
88040-900 Florian\'opolis, SC, Brazil}

\date{\today}

\begin{abstract}
We study two dimensional stripe forming systems with competing repulsive interactions decaying
as $r^{-\alpha}$. We derive an effective Hamiltonian with a short range part and a 
generalized dipolar interaction which depends on the exponent $\alpha$. 
An approximate map of this model to a known XY model with dipolar interactions allows us to conclude that,
for $\alpha <2$ long range orientational order of stripes can exist in two dimensions, and establish the universality class of the models.
When $\alpha \geq 2$ no long-range order is possible, but a phase transition in the KT universality class is still present.
These two different critical scenarios should be observed in experimentally relevant two dimensional systems like electronic liquids ($\alpha=1$) and
dipolar magnetic films ($\alpha=3$). Results from Langevin simulations of Coulomb and dipolar systems give support to the theoretical results.
\end{abstract}




\maketitle


Two dimensional isotropic systems in which a short range attractive interaction competes with a repulsive interaction 
decaying as a power law of the form $r^{-\alpha}$ have been widely studied~\cite{SeAn1995,DeSa1996,NuRuKiCh1999,GrTaVi2000,Mu2002,OrLoDi2009,
PoGoSaBiPeVi2010,BaRiSt2013}. 
These include, as physically relevant examples, 
the dipolar $(r^{-3})$ and the Coulomb $(r^{-1})$ interaction as the repulsive part of the total energy of the system.
Dipolar interactions competing with exchange and uniaxial anisotropy arise, e.g. in ultra-thin ferromagnetic films with 
perpendicular anisotropy~\cite{VaStMaPiPoPe2000,WuWoSc2004,AbVe2009} while long range Coulomb interactions appear in 
low dimensional electron systems and may be relevant to understand the low temperature phase behavior
of doped Mott insulators, two dimensional quantum Hall systems and high $T_c$ superconductors~\cite{FrKi1999,Han2001,
Borzi2007,Parker2010}.
It is well known that under certain conditions of relative strength of interactions and external parameters these systems develop 
modulated stripe-like structures in two dimensions which break space rotational symmetry, similar to classical liquid-crystal systems, 
giving rise to smectic, nematic and hexatic phases~\cite{AbKaPoSa1995,KiFrEm1998,BaMeSt2013}. This analogy, based on the $180^{\circ}$
rotational symmetry of stripe structures and elongated liquid-crystal molecules, allowed to apply well known results for liquid-crystal
systems~\cite{deGPr1998,ChLu1995} to predict the qualitative, and to some extent also quantitative phase behavior of many systems with modulated 
order parameters. Nevertheless, when it is important to understand the true nature of the thermodynamic phases, the analogy between
stripe forming systems and classical liquid-crystals should not be taken at face value. 
The basic units in liquid-crystals are elongated molecules. A given molecule typically interacts with its near neighbors and due to 
its elongated form a rotation in $180^{\circ}$ of a single molecule does not alter the energy of the system. 
On the other hand, 
the smallest relevant scale of a stripe system is the modulation length. At this scale, a basic cell can be considered as
containing a single interface and then it is a dipole of opposite densities with an average linear size equal to the modulation length.
It is important to note that such dipoles will not be, in general, elementary electric or magnetic dipoles, their character will depend 
on the nature of the density order parameter under consideration. 
Having clarified this point, in principle 
all realistic low energy configurations of the system can be built from these dipole cells. Clearly, a $180^{\circ}$ rotation
of a dipole does change the energy of the system and then cannot be considered a local symmetry. The system is only symmetric under global
rotations of $180^{\circ}$. Furthermore, when long range interactions are
present, it is well known that the behavior of the systems may be very different from those with only short range interactions, which
represent the vast majority of classical liquid-crystal systems. A study of the nature of low temperature phases of stripe forming systems
should take these elements into account.


Consider a coarse grain Hamiltonian in two dimensions of the form
\begin{eqnarray}
\nonumber
\mathcal{H}[\phi(\vec{x})]&=&\frac{1}{2}\int d^2 x\,\left(\vec{\nabla}\phi(\vec{x}) \right)^2
\nonumber \\ \nonumber
&+&\frac{1}{2}\int d^2x\int d^2x'\ \phi(\vec{x}){J}(\vec{x}-\vec{x'})\phi(\vec{x'})\\
&+& \frac{1}{2\beta}\int d^2x\ V(\phi(\vec{x})),
\label{Ham}
\end{eqnarray}
where $\beta=1/k_BT$ and $V(\phi)=-\frac{r}{2}\phi^2+\frac{u}{4}\phi^4$ 
is a local potential that could be seen as an entropic contribution and which exact form is not important to our work.
The long range repulsive interaction has the form $J(\vec{x})=J/\vert\vec{x}\vert^\alpha$ which allows to analyze in a unified way short range (large $\alpha$)
and long range (small $\alpha$) interactions. Physically relevant examples are the Coulomb interaction ($\alpha=1$) and the dipolar
interaction between out-of-plane magnetic moments ($\alpha=3$). 
It is well known that at low temperatures this kind of systems display stripe-like patterns in the form of spatial modulations of the
density $\phi(\vec{x})$~\cite{Br1975,ToNe1981,SeAn1995,BaSt2007} in a direction represented by a wave vector $\vec k_0$. Low energy excitations 
of the stripes can be described in terms of a displacement field $u(\vec{x})$ in the form $ \phi(\vec{x})=\sum_n\phi_n\cos(nk_0 x +nk_0 u(\vec{x}))$,
where $x$ is the average direction of the modulation and $k_0$ stands for the modulus of $\vec{k}_0$. If $u(\vec{x})$ varies smoothly in space it is possible to
define a local wave vector $\vec{k}_0+k_0\vec{\nabla} u(\vec{x})$.

\begin{figure}[ht!]
\includegraphics[scale=0.6]{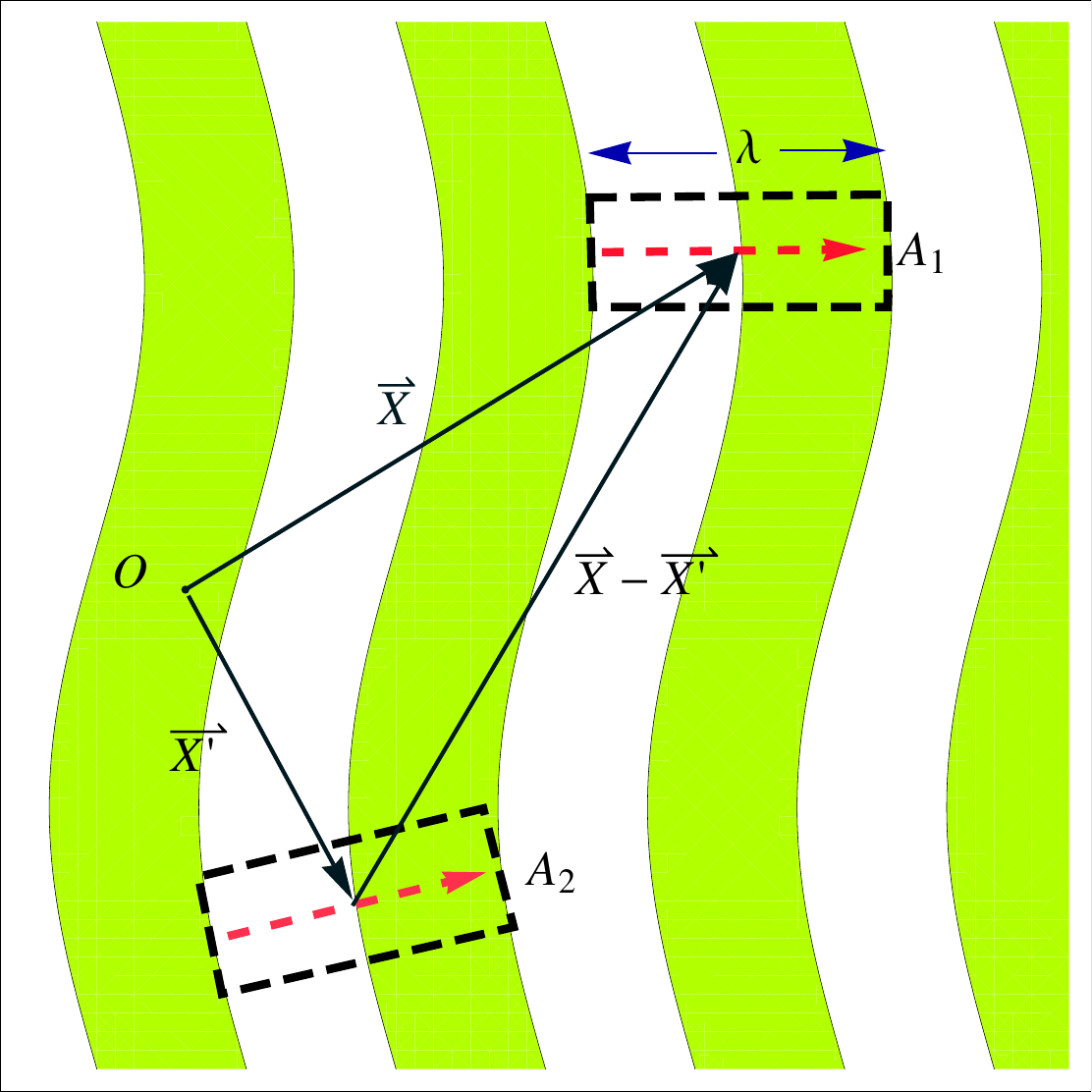}
\caption{Schematic representation of the long range interaction between two elementary stripe dipoles.}
\label{dipoles}
\end{figure}

The effective Hamiltonian (\ref{Ham}) when expressed in terms of $u(\vec x)$ has local and non-local parts $\mathcal{H}=\mathcal{H}_l+\mathcal{H}_{nl}$ 
(see Supplemental Material). 
Expanding the local component to quadratic order in the fluctuation field $u$, it can be written in Fourier space as~\cite{So1987,deGPr1998,ToNe1981}:
\begin{equation}
 \mathcal{H}_l=\mathcal{H}_{0l}+\frac{1}{2}\int\frac{d^2k}{(2\pi)^2}(\gamma_xk_x^2+\gamma_yk_y^4)\hat{u}(\vec{k})\hat{u}(-\vec{k})
\end{equation}
where $\gamma_x$ and $\gamma_y$ are elastic coefficients which are simply related to the parameters of the original Hamiltonian and
$\mathcal{H}_{0l}$ represents the local contribution to the energy for an unperturbed stripe. It is well known that this form for the local
fluctuations of the stripe pattern leads to a divergence of the mean square of the displacement field, implying the abscence of long
range positional order in the system. This is the standard situation in liquid-crystalline systems. We go on to consider the effect of 
the tail of the long range interaction in the fluctuation spectrum.
The non-local component can be taken into account properly by considering the long range interaction between a pair of stripe dipoles
as shown schematically in Figure \ref{dipoles}.
The interaction between a pair of dipoles is given by:
\begin{equation}
 \delta \mathcal{H}_{nl}=\frac{J}{2}\int_{\delta A_1} d^2x\int_{\delta A_2} d^2x'\ \frac{\phi(\vec{x})\phi(\vec{x'})}{\vert\vec{x}-\vec{x'}\vert^\alpha}
\label{lrange}
\end{equation}
where $\delta A_1$ and $\delta A_2$ are the corresponding areas (see Figure \ref{dipoles}). If $\lambda$ is the modulation length of the stripe
pattern, in the limit $\vert\vec{x}-\vec{x}'\vert\gg\lambda$ a multipolar expansion of the interaction (\ref{lrange}) leads to (see Supplemental
Material):
\begin{eqnarray}
\nonumber
\mathcal{H}_{nl}&=&\frac{1}{2}\gamma\int d^2x\int d^2x'\Omega(\vert\vec{x}-\vec{x}'\vert)\left(\frac{\vec{e}(\vec{x})\cdot\vec{e}(\vec{x}')}
{\vert\vec{x}-\vec{x}'\vert^{\alpha+2}} \right.\\
 &-&\left. (\alpha+2)\frac{\vec{e}(\vec{x})\cdot(\vec{x}-\vec{x}')\vec{e}(\vec{x}')\cdot(\vec{x}-\vec{x}')}{\vert\vec{x}-\vec{x}'\vert^{\alpha+4}}\right).
\label{Hnl}
\end{eqnarray}
In this expression $\gamma=\alpha J{P^2}$ and $P=\frac{1}{\lambda}\int_{\lambda}dx  x\phi(x)$ is the modulus of the dipolar moment.
The unit vectors $\vec{e}(\vec x)$ give the orientation of the dipoles which point along the local wave vector of the stripe pattern and $\Omega(x)$
is a short range cutoff. Here we have neglected fluctuations in the modulation length and accordingly the elastic coefficient $\gamma$ is evaluated 
in its mean field value (see Supplemental Material for a discussion on relevant fluctuations). As we can see from the expression obtained, the 
long range repulsive interaction is responsible for a generalized dipolar contribution to the total energy.

Considering again small fluctuations in the direction of the wave vector $\vec k_0$, we can write
$\vec{e}(\vec{x})\approx\frac{\vec{k}_0}{k_0}+\vec{\nabla}u(\vec{x})$,
which leads (considering that $\vec{k}_0$ points in $x$ direction) to:
\begin{eqnarray}
\nonumber
\Delta\mathcal{H}_{nl}&=&\frac{1}{2}\gamma\int d^2x\int d^2x'\Omega(\vert\vec{x}-\vec{x}'\vert)\left(\frac{\partial_yu(\vec{x})\partial_{y'}u(\vec{x}')}
{\vert\vec{x}-\vec{x}'\vert^{\alpha+2}} \right.\\
&-&\left. (\alpha+2)\frac{(y-y')^2\partial_yu(\vec{x})\partial_{y'}u(\vec{x}')}{\vert\vec{x}-\vec{x}'\vert^{\alpha+4}}\right).
\end{eqnarray}
Thus, the effective Hamiltonian for the displacement field $u(\vec{x})$ results: 
\begin{equation}
\Delta\mathcal{H}=
\frac{1}{2}\int\frac{d^2k}{(2\pi)^2}(\gamma_xk_x^2+\gamma_yk_y^4+\gamma_{nl}k^{\alpha-2}{k_y}^4)\hat{u}(\vec{k})\hat{u}(-\vec{k})
\label{Heff}
\end{equation}
where $\gamma_{nl}=\gamma\alpha(2+\alpha)^2C(\alpha+4)$ and the function $C(\alpha)=2^{2-\alpha}{\Gamma(\frac{2-\alpha}{2})}/{\Gamma(\frac{\alpha}{2})}$
with $\Gamma(x)$ being the Gamma function.
From the previous considerations we are now in a position to analyze the stability of positional and orientational order of the
stripe structures when the long range interactions are taken into account.

{\em Positional order}: 
From the effective Hamiltonian (\ref{Heff}) we can see that for $\alpha \geq 2$ the $\gamma_yk_y^4$ term dominates over 
$\gamma_{nl} k^{\alpha-2}{k_y}^4$
in the long wavelength limit, i.e. for sufficiently short range interaction no positional order is possible. 
If $\alpha<2$ the term $\gamma_{nl} k^{\alpha-2}{k_y}^4$ 
dominates over $\gamma_yk_y^4$, but even in this longe range interacting regime it is easy to check that the average square 
fluctuations $\langle u^2\rangle=k_BT\int\frac{d^2k}{(2\pi)^2}\left(\gamma_xk_x^2+\gamma_yk_y^4+\gamma_{nl} k^{\alpha-2}{k_y}^4\right)^{-1}$ 
diverge with some power of the system size for any $\alpha>0$.

{\em Orientational order}:
It is well known that in systems with short range interactions orientational order can be weakened by the presence of topological defects
~\cite{KoTh1973,ToNe1981}. The typical situation in two dimensional systems with continuous symmetries is that only quasi-long range order
is possible when interactions are of sufficiently short range~\cite{MeWa1966}. Nevertheless it is commonly argued that even in systems with long 
range interactions, like Coulomb or dipolar interactions, shielding effects make the effective interactions short ranged.  
Here we revisit this question, considering explicitly the effects of the range of the interactions and show that, although the shielding occurs, 
the effective interactions are still capable of stabilizing a long-range-ordered nematic phase in two dimensions for long enough 
interaction range.

At low temperatures the stripe structure can be thought of as composed by a mosaic of domains of average size $\xi_u$ corresponding to the correlation
length of the displacement field $u(\vec{x})$. The orientation of each domain is a natural order parameter which can be
described by a unit vector $\vec n$. This vector represents the mean orientation of the elementary dipoles inside a domain and consequently 
it is defined in terms of the unit vectors $\vec{e}(\vec{x})$ previously defined in (\ref{Hnl}) as :
\begin{equation}
\vec{n}(\vec{x})=\frac{\int_{\Delta A_u} d^2x\,\vec{e}(\vec{x})}{\left\vert\int_{\Delta A_u} d^2x\,\vec{e}(\vec{x})\right\vert},
\end{equation}
where $\Delta A_u$ is the area of the domain, and it is over this area that a coarse graining process is made. 
Proceeding as in the analysis of positional order, we can separate the contribution to the orientational energy into two parts, a local part 
coming from interactions between
nearby domains and a non-local one due to interactions between far apart domains. In the long wavelength limit, at the scale of the correlation length 
$\xi_u$, the effective interaction between nearby domains will be of the form:
\begin{equation}
\Delta\mathcal{H}_{ol}=\frac{\gamma_{ol}}{2}\int d^2x\, (\vec{\nabla}\theta)^2(\vec{x}) 
\label{Hol}
\end{equation}
where $\theta(\vec{x})$ is the angle between two neighboring domains pointing along directions $\vec n$ and $\vec n'$. The elastic coefficient 
$\gamma_{ol}$ can be estimated to be $J\alpha^2 P^2/(4\xi_u^{\alpha-2})$. To continue with our analysis we realize that over length 
of order $\xi_u$, deviations of the local directors $\vec{e}(x)$ are small. This means after a coarse graining process, the interactions between 
far apart well polarized domains ( of typical size $\xi_u\times\xi_u$ ) has the same form of Eq. (\ref{Hnl}):
\begin{eqnarray}
\nonumber
\mathcal{H}_{onl}&=&\frac{\gamma}{2}\int d^2x\int d^2x'\,\Omega(\vert\vec{x}-\vec{x}'\vert)\left(\frac{\vec{n}(\vec{x})\cdot\vec{n}(\vec{x}')}
{\vert\vec{x}-\vec{x}'\vert^{\alpha+2}} \right.\\
 &-&\left. (\alpha+2)\frac{\vec{n}(\vec{x})\cdot(\vec{x}-\vec{x}')\vec{n}(\vec{x}')\cdot(\vec{x}-\vec{x}')}{\vert\vec{x}-\vec{x}'\vert^{\alpha+4}}\right)
\label{Honl}
\end{eqnarray}
as a consequence of the principle of superposition. Then, $\mathcal{H}_{o}=\mathcal{H}_{ol}+\mathcal{H}_{onl}$ is the complete orientational effective Hamiltonian. 
This is one of the main results of
our work (see Supplemental Material). Note that usually the effective orientational energy is taken to be composed only by the local part, corresponding to smooth variations
in the mean directions of neighboring striped domains. We will see in the sequel that the presence of the second (non-local) term can potentially
change the universality class of the orientational order in the system. A renormalization group study of the orientational effective Hamiltonian 
$\mathcal{H}_o$ has been done
before in Ref. \onlinecite{MaSc2004} for the case $\alpha=1$, which corresponds to a dipolar XY model. In that reference, the authors were able to
renormalize the model and, importantly, they showed that the universal properties are not changed
by the presence of the anisotropic part of the interaction. Furthermore, they showed that a whole family of models with isotropic long range
interactions of the form $\int_{\vec k} |k|^{\sigma}\vec{S}(\vec k)\vec{S}(-\vec k)$ behave in qualitatively the same way as the dipolar XY model as
long as the range $\sigma <2$. Once the mapping between these models and ours is established then the critical properties of the stripe
forming systems are known. In fact, the Fourier transform of the isotropic term in Eq. (\ref{Honl}) is proportional to
$\int_{\vec k} |k|^{\alpha}\vec{n}(\vec k)\vec{n}(-\vec k)$ \cite{BaRiSt2013}. Then, one immediately see that for $\alpha \geq 2$ the leading term
in $\mathcal{H}_o$ is quadratic in $k$. In this case the low temperature physics of the system is that of the two dimensional short range XY
model, i.e. there is a phase transition of the Kosterlitz-Thouless type at a critical temperature $T_{KT}$. In a system with dipolar interactions
$\alpha=3$ and then we expect it to have an isotropic-nematic phase transition of the KT type, as anticipated in previous works based on analysis
of fluctuations of the local part of the effective Hamiltonian \cite{BaSt2007,BaSt2009}. In this case nematic order is quasi-long-range with
algebraically decaying correlations. However, when $\alpha <2$ the physics changes according
to the results of Ref. \onlinecite{MaSc2004}. Now, the non-local part in $\mathcal{H}_o$ is relevant and rules the low temperature phase transition.
In fact, the long range nature of the interactions in this sector are able to stabilize a nematic phase with truly long range order below a
critical temperature $T_c$. It is possible to show, in the framework of renormalization group equations, that the critical properties of the
systems for $\alpha <2$ show some peculiar characteristics, for example \cite{MaSc2004}:
\begin{itemize}
 \item in the critical region, the correlation length diverges exponentially at $T_c$, from both sides, as 
 $\xi_o \propto\exp\left(\frac{b}{\sqrt{\vert T_c-T\vert}}\right)$,
reminiscent of the KT transition behavior.
 \item For $T<T_c$ in the critical region, the average dipolar moment behaves as $M\propto\xi_o(T)^{-(2-\alpha)/2}$, showing the existence 
of long range order when $\alpha <2$. 
 \item The orientational susceptibility diverges as $\chi_o\propto\xi_o(T)^{\alpha}$ in the critical region.
\end{itemize}
This kind of behavior should be observable, e.g. in systems with long range Coulomb interactions for which $\alpha=1$. This case maps onto
the dipolar XY model analyzed in \onlinecite{MaSc2004} and the results may be relevant to understand the phase behavior of two dimensional electron
systems. In the next section we show results from computer simulations of systems with $\alpha=1$ (Coulomb) and $\alpha=3$ (dipolar)
which give support to the different scenarios in both systems as described before.

{\em Simulation results}:
We performed Langevin simulations of the Hamiltonian (\ref{Ham}). The relaxational (overdamped) Langevin dynamics of the density $\phi(\vec x)$ 
is defined in reciprocal space by:
\begin{equation}
 \frac{\partial \phi}{\partial t}(\vec{k},t)= -\mathcal{A_{\alpha}}(k)\phi(\vec{k},t) -u[\phi^3]_F(\vec{k},t)+\eta(\vec{k},t)
\label{lank}
\end{equation}
where $\mathcal{A_{\alpha}}(k)$ is the spectrum of fluctuations, i.e. the Fourier transform of the quadratic part of the effective Hamiltonian (\ref{Ham}),
  $[\phi^3]_F(\vec{k},t)$ stands for the Fourier transform of
$\phi^3(\vec{x},t)$ and $\eta(\vec{k},t)$ represents a Gaussian white noise
with correlations $<\eta(\vec{k},t)\eta(\vec{k}',t')>=(2\pi)^2 2T \delta(\vec{k}+\vec{k}')\delta(t-t')$, where $T$ is the effective temperature 
of the heat bath.  We worked with two forms of $\mathcal{A_{\alpha}}(k)$, the first one
$\mathcal{A}_3(k) = a_2 (k - k_0)^2 - r$ encodes the linear dependence
of the isotropic dipolar interaction with $k$, with $a_2$ and $r$ constants. The second form is
$\mathcal{A}_1(k) = a_2 (k^2 + 2 k_0^3/k -3k_0^2) - r$,
corresponding to the Coulomb interaction proportional to $1/k$ in two dimensional Fourier space. The parameters were chosen
such as to have the same values of $\mathcal{A_{\alpha}}(k)$ close to
the minimum at $k_0$. To ensure this we have set $a_2=1$ for the dipolar 
and $a_2=1/3$ for the Coulomb cases. In both cases we set $r=1$ and $k_0 = 1$.

For the numerical simulations we have used an implicit first-order
scheme for the numerical integration of (\ref{lank}) in the Fourier
space, a procedure that guarantees good numerical stability with time
step $dt=0.1$, as established in previous works
\cite{NiSt2007,DiMeMuNiSt2011}. In the adimensional form, the
periodicity of the stripes are set by the lattice constant $dx$ of a
2d square grid with linear size $L=MN$, so that
$\vec{k}=(k_x/dx,k_y/dx)$ with $k_i = 2\pi n_i/L$ and
$dx=\pi/M$. Within this scheme, the stripe length span $M$ lattice
sites and the linear system size is such that contains $N$ stripes. We
fixed $M=11$ in order to have smooth domain walls.

After an estimation of the equilibration and correlation times from
high temperature quenches, we performed slow cooling
experiments and found that below $T\simeq 0.59$ ($T\simeq 0.45$) the
dipolar (Coulomb)
systems find themselves in the low temperature phases (with
orientational order) for all system sizes. 
Above those temperatures the configurations are in a state
usually called {\it liquid of stripes}, where both positional and
orientational correlation lengths are finite. So we concentrated on
equilibrium simulations for $T=0.57$ ($T=0.43$) for system sizes
ranging from $(12\times11)^2$ up to $(66\times11)^2$. 
The orientational order was quantified through the local director field
$\vec{v}(\vec{x}) =\vec{\nabla }\phi (\vec{x})/|\vec {\nabla }\phi
(\vec{x})|$ by measuring $Q=\langle \cos{2\theta(\vec{x})}\rangle$
 and its corresponding susceptibility $\chi_o$, where
$\theta(\vec{x})$ is the angle defining the local orientation of the director
field. 

\begin{figure}[ht!]
\includegraphics[scale=0.7]{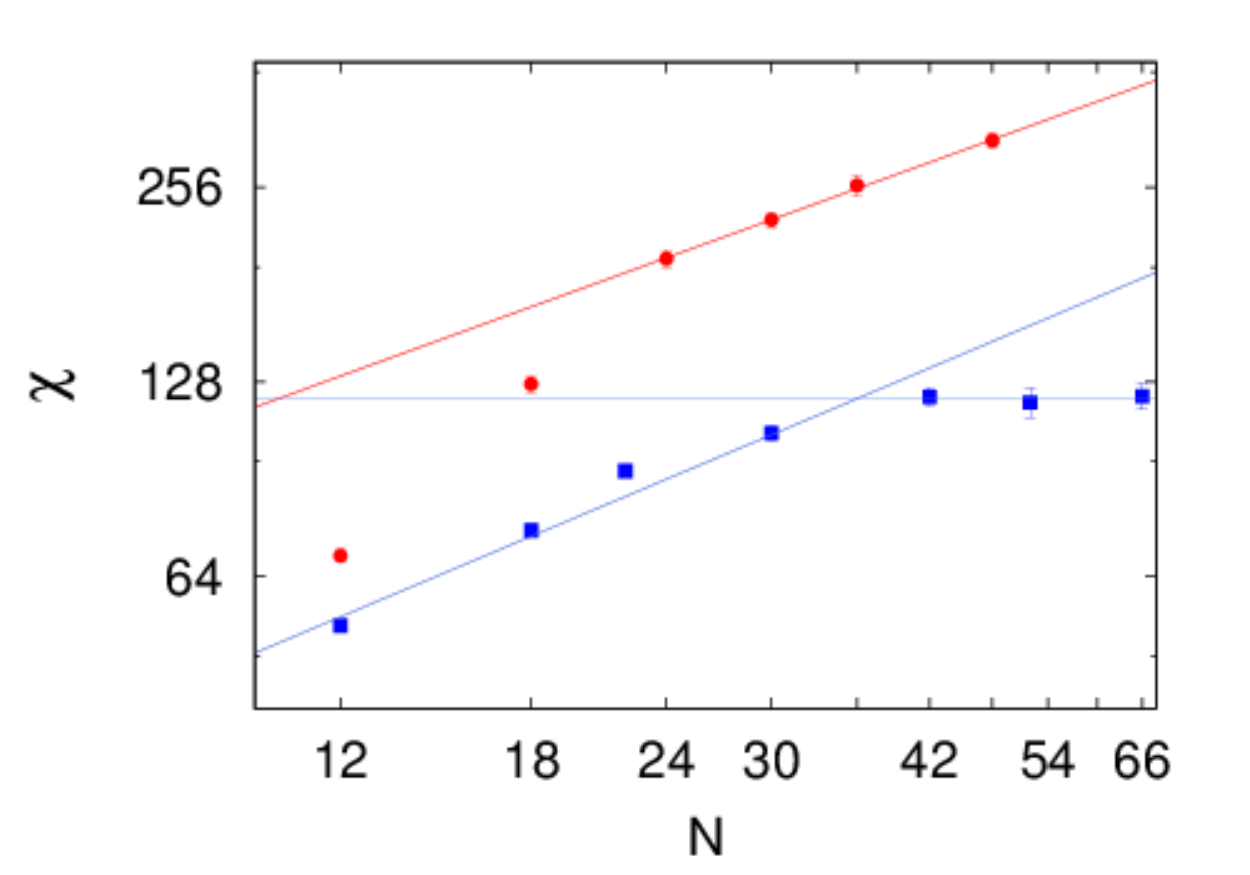}
\caption{Orientational susceptibility of the Coulomb (blue squares) and dipolar (red dots) models as function of the linear size of the 
systems in log-log scale. The full lines are power law fits with exponents 0.7 (blue) and 0.6 (red).}
\label{susc}
\end{figure}

The previous analysis implies that, in the limit of large system
sizes, the orientational susceptibility for interactions with
$\alpha=1$ and $\alpha=3$ should be qualitatively different
for $T<T_c$.  In the Coulomb case, the second order nature of the phase transition
should imply that the susceptibility must be finite when $N \to \infty$.  
On the other hand, for dipolar interactions the transition should
be of the KT type, implying a monotonic (logarithmic) increase of $\chi_o$
with system size, which should diverge in the thermodynamic limit for all $T\leq T_{KT}$.
Results for the orientational susceptibility as a function
of the linear system size ($N=L/11$) from simulations are shown in Fig.
\ref{susc} for the two characteristic temperatures cited above, corresponding to the low temperature 
phase of each model. Although computational limitations prevent us to reach very large system sizes,
it is clarly observed that
the susceptibility of the Coulomb system ($\alpha=1$) first grows with $N$ but eventually suffers a
crossover and then saturates at a fixed value for the largest sizes. On the
other hand, the susceptibility in the dipolar system ($\alpha=3$)
shows a power law increase with system
size, a behavior consistent with that of a KT-like critical phase. Of course, we cannot conclude that
$\chi_o$ will not saturate at larger $N$'s, but the different trend observed in both systems for
equivalent parameter values is a strong indication that the theoretical results are indeed correct.

In summary, we have shown that two-dimensional stripe forming
systems with isotropic competing interactions can be classified into
two universality classes: for sufficiently short-range interactions a
Kosterlitz-Thouless transition from an isotropic to a quasi-long-range orientational order
phase takes place with the well known phenomenology of defect-mediated phase
transitions; but, for sufficiently long-range repulsive interactions a second
order phase transition with some unusual characteristics drives the system from
the isotropic to a fully long-range orientational order phase. These results 
improve considerably the understanding of the nature of phase transtions in stripe
forming systems and may be relevant to a wide variety of systems, particularly
the strong correlated regime of two dimensional ``electronic liquid-crystals'' phases
and modulated phases in ultrathin magnetic films with perpendicular anisotropy.



{\em Acknowledgments}: We gratefully acknowledge partial financial
support from CNPq (Brazil) and the {\it Laborat\'orio de F\'isica
  Computacional} from IF-UFRGS for the use of the cluster Ada.


\begin{thebibliography}{30}%
\makeatletter
\providecommand \@ifxundefined [1]{%
 \@ifx{#1\undefined}
}%
\providecommand \@ifnum [1]{%
 \ifnum #1\expandafter \@firstoftwo
 \else \expandafter \@secondoftwo
 \fi
}%
\providecommand \@ifx [1]{%
 \ifx #1\expandafter \@firstoftwo
 \else \expandafter \@secondoftwo
 \fi
}%
\providecommand \natexlab [1]{#1}%
\providecommand \enquote  [1]{``#1''}%
\providecommand \bibnamefont  [1]{#1}%
\providecommand \bibfnamefont [1]{#1}%
\providecommand \citenamefont [1]{#1}%
\providecommand \href@noop [0]{\@secondoftwo}%
\providecommand \href [0]{\begingroup \@sanitize@url \@href}%
\providecommand \@href[1]{\@@startlink{#1}\@@href}%
\providecommand \@@href[1]{\endgroup#1\@@endlink}%
\providecommand \@sanitize@url [0]{\catcode `\\12\catcode `\$12\catcode
  `\&12\catcode `\#12\catcode `\^12\catcode `\_12\catcode `\%12\relax}%
\providecommand \@@startlink[1]{}%
\providecommand \@@endlink[0]{}%
\providecommand \url  [0]{\begingroup\@sanitize@url \@url }%
\providecommand \@url [1]{\endgroup\@href {#1}{\urlprefix }}%
\providecommand \urlprefix  [0]{URL }%
\providecommand \Eprint [0]{\href }%
\providecommand \doibase [0]{http://dx.doi.org/}%
\providecommand \selectlanguage [0]{\@gobble}%
\providecommand \bibinfo  [0]{\@secondoftwo}%
\providecommand \bibfield  [0]{\@secondoftwo}%
\providecommand \translation [1]{[#1]}%
\providecommand \BibitemOpen [0]{}%
\providecommand \bibitemStop [0]{}%
\providecommand \bibitemNoStop [0]{.\EOS\space}%
\providecommand \EOS [0]{\spacefactor3000\relax}%
\providecommand \BibitemShut  [1]{\csname bibitem#1\endcsname}%
\let\auto@bib@innerbib\@empty
\bibitem [{\citenamefont {Seul}\ and\ \citenamefont
  {Andelman}(1995)}]{SeAn1995}%
  \BibitemOpen
  \bibfield  {author} {\bibinfo {author} {\bibfnamefont {M.}~\bibnamefont
  {Seul}}\ and\ \bibinfo {author} {\bibfnamefont {D.}~\bibnamefont
  {Andelman}},\ }\href@noop {} {\bibfield  {journal} {\bibinfo  {journal}
  {Science}\ }\textbf {\bibinfo {volume} {267}},\ \bibinfo {pages} {476}
  (\bibinfo {year} {1995})}\BibitemShut {NoStop}%
\bibitem [{\citenamefont {Deutsch}\ and\ \citenamefont
  {Safran}(1996)}]{DeSa1996}%
  \BibitemOpen
  \bibfield  {author} {\bibinfo {author} {\bibfnamefont {A.}~\bibnamefont
  {Deutsch}}\ and\ \bibinfo {author} {\bibfnamefont {S.~A.}\ \bibnamefont
  {Safran}},\ }\href {\doibase 10.1103/PhysRevE.54.3906} {\bibfield  {journal}
  {\bibinfo  {journal} {Phys. Rev. E}\ }\textbf {\bibinfo {volume} {54}},\
  \bibinfo {pages} {3906} (\bibinfo {year} {1996})}\BibitemShut {NoStop}%
\bibitem [{\citenamefont {Nussinov}\ \emph {et~al.}(1999)\citenamefont
  {Nussinov}, \citenamefont {Rudnick}, \citenamefont {Kivelson},\ and\
  \citenamefont {Chayes}}]{NuRuKiCh1999}%
  \BibitemOpen
  \bibfield  {author} {\bibinfo {author} {\bibfnamefont {Z.}~\bibnamefont
  {Nussinov}}, \bibinfo {author} {\bibfnamefont {J.}~\bibnamefont {Rudnick}},
  \bibinfo {author} {\bibfnamefont {S.~A.}\ \bibnamefont {Kivelson}}, \ and\
  \bibinfo {author} {\bibfnamefont {L.~N.}\ \bibnamefont {Chayes}},\ }\href
  {\doibase 10.1103/PhysRevLett.83.472} {\bibfield  {journal} {\bibinfo
  {journal} {Phys. Rev. Lett.}\ }\textbf {\bibinfo {volume} {83}},\ \bibinfo
  {pages} {472} (\bibinfo {year} {1999})}\BibitemShut {NoStop}%
\bibitem [{\citenamefont {Grousson}\ \emph {et~al.}(2000)\citenamefont
  {Grousson}, \citenamefont {Tarjus},\ and\ \citenamefont {Viot}}]{GrTaVi2000}%
  \BibitemOpen
  \bibfield  {author} {\bibinfo {author} {\bibfnamefont {M.}~\bibnamefont
  {Grousson}}, \bibinfo {author} {\bibfnamefont {G.}~\bibnamefont {Tarjus}}, \
  and\ \bibinfo {author} {\bibfnamefont {P.}~\bibnamefont {Viot}},\ }\href
  {\doibase 10.1103/PhysRevE.62.7781} {\bibfield  {journal} {\bibinfo
  {journal} {Phys. Rev. E}\ }\textbf {\bibinfo {volume} {62}},\ \bibinfo
  {pages} {7781} (\bibinfo {year} {2000})}\BibitemShut {NoStop}%
\bibitem [{\citenamefont {Muratov}(2002)}]{Mu2002}%
  \BibitemOpen
  \bibfield  {author} {\bibinfo {author} {\bibfnamefont {C.~B.}\ \bibnamefont
  {Muratov}},\ }\href {\doibase 10.1103/PhysRevE.66.066108} {\bibfield
  {journal} {\bibinfo  {journal} {Phys. Rev. E}\ }\textbf {\bibinfo {volume}
  {66}},\ \bibinfo {pages} {066108} (\bibinfo {year} {2002})}\BibitemShut
  {NoStop}%
\bibitem [{\citenamefont {Ortix}\ \emph {et~al.}(2009)\citenamefont {Ortix},
  \citenamefont {Lorenzana},\ and\ \citenamefont {Castro}}]{OrLoDi2009}%
  \BibitemOpen
  \bibfield  {author} {\bibinfo {author} {\bibfnamefont {C.}~\bibnamefont
  {Ortix}}, \bibinfo {author} {\bibfnamefont {J.}~\bibnamefont {Lorenzana}}, \
  and\ \bibinfo {author} {\bibfnamefont {C.~D.}\ \bibnamefont {Castro}},\
  }\href {\doibase http://dx.doi.org/10.1016/j.physb.2008.11.045} {\bibfield
  {journal} {\bibinfo  {journal} {Physica B: Condensed Matter}\ }\textbf
  {\bibinfo {volume} {404}},\ \bibinfo {pages} {499 } (\bibinfo {year}
  {2009})}\BibitemShut {NoStop}%
\bibitem [{\citenamefont {Portmann}\ \emph {et~al.}(2010)\citenamefont
  {Portmann}, \citenamefont {G\"olzer}, \citenamefont {Saratz}, \citenamefont
  {Billoni}, \citenamefont {Pescia},\ and\ \citenamefont
  {Vindigni}}]{PoGoSaBiPeVi2010}%
  \BibitemOpen
  \bibfield  {author} {\bibinfo {author} {\bibfnamefont {O.}~\bibnamefont
  {Portmann}}, \bibinfo {author} {\bibfnamefont {A.}~\bibnamefont {G\"olzer}},
  \bibinfo {author} {\bibfnamefont {N.}~\bibnamefont {Saratz}}, \bibinfo
  {author} {\bibfnamefont {O.~V.}\ \bibnamefont {Billoni}}, \bibinfo {author}
  {\bibfnamefont {D.}~\bibnamefont {Pescia}}, \ and\ \bibinfo {author}
  {\bibfnamefont {A.}~\bibnamefont {Vindigni}},\ }\href {\doibase
  10.1103/PhysRevB.82.184409} {\bibfield  {journal} {\bibinfo  {journal} {Phys.
  Rev. B}\ }\textbf {\bibinfo {volume} {82}},\ \bibinfo {pages} {184409}
  (\bibinfo {year} {2010})}\BibitemShut {NoStop}%
\bibitem [{\citenamefont {Barci}\ \emph
  {et~al.}(2013{\natexlab{a}})\citenamefont {Barci}, \citenamefont {Ribeiro},\
  and\ \citenamefont {Stariolo}}]{BaRiSt2013}%
  \BibitemOpen
  \bibfield  {author} {\bibinfo {author} {\bibfnamefont {D.~G.}\ \bibnamefont
  {Barci}}, \bibinfo {author} {\bibfnamefont {L.}~\bibnamefont {Ribeiro}}, \
  and\ \bibinfo {author} {\bibfnamefont {D.~A.}\ \bibnamefont {Stariolo}},\
  }\href {\doibase 10.1103/PhysRevE.87.062119} {\bibfield  {journal} {\bibinfo
  {journal} {Phys. Rev. E}\ }\textbf {\bibinfo {volume} {87}},\ \bibinfo
  {pages} {062119} (\bibinfo {year} {2013}{\natexlab{a}})}\BibitemShut
  {NoStop}%
\bibitem [{\citenamefont {Vaterlaus}\ \emph {et~al.}(2000)\citenamefont
  {Vaterlaus}, \citenamefont {Stamm}, \citenamefont {Maier}, \citenamefont
  {Pini}, \citenamefont {Politi},\ and\ \citenamefont
  {Pescia}}]{VaStMaPiPoPe2000}%
  \BibitemOpen
  \bibfield  {author} {\bibinfo {author} {\bibfnamefont {A.}~\bibnamefont
  {Vaterlaus}}, \bibinfo {author} {\bibfnamefont {C.}~\bibnamefont {Stamm}},
  \bibinfo {author} {\bibfnamefont {U.}~\bibnamefont {Maier}}, \bibinfo
  {author} {\bibfnamefont {M.~G.}\ \bibnamefont {Pini}}, \bibinfo {author}
  {\bibfnamefont {P.}~\bibnamefont {Politi}}, \ and\ \bibinfo {author}
  {\bibfnamefont {D.}~\bibnamefont {Pescia}},\ }\href@noop {} {\bibfield
  {journal} {\bibinfo  {journal} {Phys. Rev. Lett.}\ }\textbf {\bibinfo
  {volume} {84}},\ \bibinfo {pages} {2247} (\bibinfo {year}
  {2000})}\BibitemShut {NoStop}%
\bibitem [{\citenamefont {Wu}\ \emph {et~al.}(2004)\citenamefont {Wu},
  \citenamefont {Won}, \citenamefont {Scholl}, \citenamefont {Doran},
  \citenamefont {Zhao}, \citenamefont {Jin},\ and\ \citenamefont
  {Qiu}}]{WuWoSc2004}%
  \BibitemOpen
  \bibfield  {author} {\bibinfo {author} {\bibfnamefont {Y.~Z.}\ \bibnamefont
  {Wu}}, \bibinfo {author} {\bibfnamefont {C.}~\bibnamefont {Won}}, \bibinfo
  {author} {\bibfnamefont {A.}~\bibnamefont {Scholl}}, \bibinfo {author}
  {\bibfnamefont {A.}~\bibnamefont {Doran}}, \bibinfo {author} {\bibfnamefont
  {H.~W.}\ \bibnamefont {Zhao}}, \bibinfo {author} {\bibfnamefont {X.~F.}\
  \bibnamefont {Jin}}, \ and\ \bibinfo {author} {\bibfnamefont {Z.~Q.}\
  \bibnamefont {Qiu}},\ }\href@noop {} {\bibfield  {journal} {\bibinfo
  {journal} {Physical Review Letters}\ }\textbf {\bibinfo {volume} {93}},\
  \bibinfo {pages} {117205} (\bibinfo {year} {2004})}\BibitemShut {NoStop}%
\bibitem [{\citenamefont {Abu-Libdeh}\ and\ \citenamefont
  {Venus}(2009)}]{AbVe2009}%
  \BibitemOpen
  \bibfield  {author} {\bibinfo {author} {\bibfnamefont {N.}~\bibnamefont
  {Abu-Libdeh}}\ and\ \bibinfo {author} {\bibfnamefont {D.}~\bibnamefont
  {Venus}},\ }\href@noop {} {\bibfield  {journal} {\bibinfo  {journal} {Phys.
  Rev. B}\ }\textbf {\bibinfo {volume} {80}},\ \bibinfo {pages} {184412}
  (\bibinfo {year} {2009})}\BibitemShut {NoStop}%
\bibitem [{\citenamefont {Fradkin}\ and\ \citenamefont
  {Kivelson}(1999)}]{FrKi1999}%
  \BibitemOpen
  \bibfield  {author} {\bibinfo {author} {\bibfnamefont {E.}~\bibnamefont
  {Fradkin}}\ and\ \bibinfo {author} {\bibfnamefont {S.~A.}\ \bibnamefont
  {Kivelson}},\ }\href {\doibase 10.1103/PhysRevB.59.8065} {\bibfield
  {journal} {\bibinfo  {journal} {Phys. Rev. B}\ }\textbf {\bibinfo {volume}
  {59}},\ \bibinfo {pages} {8065} (\bibinfo {year} {1999})}\BibitemShut
  {NoStop}%
\bibitem [{\citenamefont {Han}\ \emph {et~al.}(2001)\citenamefont {Han},
  \citenamefont {Wang},\ and\ \citenamefont {Lee}}]{Han2001}%
  \BibitemOpen
  \bibfield  {author} {\bibinfo {author} {\bibfnamefont {J.}~\bibnamefont
  {Han}}, \bibinfo {author} {\bibfnamefont {Q.-H.}\ \bibnamefont {Wang}}, \
  and\ \bibinfo {author} {\bibfnamefont {D.-H.}\ \bibnamefont {Lee}},\ }\href
  {\doibase 10.1142/S021797920100468X} {\bibfield  {journal} {\bibinfo
  {journal} {Int. J. Mod. Phys. B}\ }\textbf {\bibinfo {volume} {15}},\
  \bibinfo {pages} {1117} (\bibinfo {year} {2001})}\BibitemShut {NoStop}%
\bibitem [{\citenamefont {Borzi}\ \emph {et~al.}(2007)\citenamefont {Borzi},
  \citenamefont {Grigera}, \citenamefont {Farrell}, \citenamefont {Perry},
  \citenamefont {Lister}, \citenamefont {Lee}, \citenamefont {Tennant},
  \citenamefont {Maeno},\ and\ \citenamefont {Mackenzie}}]{Borzi2007}%
  \BibitemOpen
  \bibfield  {author} {\bibinfo {author} {\bibfnamefont {R.~A.}\ \bibnamefont
  {Borzi}}, \bibinfo {author} {\bibfnamefont {S.~A.}\ \bibnamefont {Grigera}},
  \bibinfo {author} {\bibfnamefont {J.}~\bibnamefont {Farrell}}, \bibinfo
  {author} {\bibfnamefont {R.~S.}\ \bibnamefont {Perry}}, \bibinfo {author}
  {\bibfnamefont {S.~J.~S.}\ \bibnamefont {Lister}}, \bibinfo {author}
  {\bibfnamefont {S.~L.}\ \bibnamefont {Lee}}, \bibinfo {author} {\bibfnamefont
  {D.~A.}\ \bibnamefont {Tennant}}, \bibinfo {author} {\bibfnamefont
  {Y.}~\bibnamefont {Maeno}}, \ and\ \bibinfo {author} {\bibfnamefont {A.~P.}\
  \bibnamefont {Mackenzie}},\ }\href {\doibase 10.1126/science.1134796}
  {\bibfield  {journal} {\bibinfo  {journal} {Science}\ }\textbf {\bibinfo
  {volume} {315}},\ \bibinfo {pages} {214} (\bibinfo {year}
  {2007})}\BibitemShut {NoStop}%
\bibitem [{\citenamefont {Parker}\ \emph {et~al.}(2010)\citenamefont {Parker},
  \citenamefont {Aynajian}, \citenamefont {da~Silva~Neto}, \citenamefont
  {Pushp}, \citenamefont {Ono}, \citenamefont {Wen}, \citenamefont {Xu},
  \citenamefont {Gu},\ and\ \citenamefont {Yazdani}}]{Parker2010}%
  \BibitemOpen
  \bibfield  {author} {\bibinfo {author} {\bibfnamefont {C.~V.}\ \bibnamefont
  {Parker}}, \bibinfo {author} {\bibfnamefont {P.}~\bibnamefont {Aynajian}},
  \bibinfo {author} {\bibfnamefont {E.~H.}\ \bibnamefont {da~Silva~Neto}},
  \bibinfo {author} {\bibfnamefont {A.}~\bibnamefont {Pushp}}, \bibinfo
  {author} {\bibfnamefont {S.}~\bibnamefont {Ono}}, \bibinfo {author}
  {\bibfnamefont {J.}~\bibnamefont {Wen}}, \bibinfo {author} {\bibfnamefont
  {Z.}~\bibnamefont {Xu}}, \bibinfo {author} {\bibfnamefont {G.}~\bibnamefont
  {Gu}}, \ and\ \bibinfo {author} {\bibfnamefont {A.}~\bibnamefont {Yazdani}},\
  }\href {http://dx.doi.org/10.1038/nature09597} {\bibfield  {journal}
  {\bibinfo  {journal} {Nature}\ }\textbf {\bibinfo {volume} {468}},\ \bibinfo
  {pages} {677} (\bibinfo {year} {2010})}\BibitemShut {NoStop}%
\bibitem [{\citenamefont {Abanov}\ \emph {et~al.}(1995)\citenamefont {Abanov},
  \citenamefont {Kalatsky}, \citenamefont {Pokrovsky},\ and\ \citenamefont
  {Saslow}}]{AbKaPoSa1995}%
  \BibitemOpen
  \bibfield  {author} {\bibinfo {author} {\bibfnamefont {A.}~\bibnamefont
  {Abanov}}, \bibinfo {author} {\bibfnamefont {V.}~\bibnamefont {Kalatsky}},
  \bibinfo {author} {\bibfnamefont {V.~L.}\ \bibnamefont {Pokrovsky}}, \ and\
  \bibinfo {author} {\bibfnamefont {W.~M.}\ \bibnamefont {Saslow}},\
  }\href@noop {} {\bibfield  {journal} {\bibinfo  {journal} {Phys. Rev. B}\
  }\textbf {\bibinfo {volume} {51}},\ \bibinfo {pages} {1023} (\bibinfo {year}
  {1995})}\BibitemShut {NoStop}%
\bibitem [{\citenamefont {Kivelson}\ \emph {et~al.}(1998)\citenamefont
  {Kivelson}, \citenamefont {Fradkin},\ and\ \citenamefont
  {Emery}}]{KiFrEm1998}%
  \BibitemOpen
  \bibfield  {author} {\bibinfo {author} {\bibfnamefont {S.~A.}\ \bibnamefont
  {Kivelson}}, \bibinfo {author} {\bibfnamefont {E.}~\bibnamefont {Fradkin}}, \
  and\ \bibinfo {author} {\bibfnamefont {V.~J.}\ \bibnamefont {Emery}},\
  }\href@noop {} {\bibfield  {journal} {\bibinfo  {journal} {Nature}\ }\textbf
  {\bibinfo {volume} {393}},\ \bibinfo {pages} {550} (\bibinfo {year}
  {1998})}\BibitemShut {NoStop}%
\bibitem [{\citenamefont {Barci}\ \emph
  {et~al.}(2013{\natexlab{b}})\citenamefont {Barci}, \citenamefont
  {Mendoza-Coto},\ and\ \citenamefont {Stariolo}}]{BaMeSt2013}%
  \BibitemOpen
  \bibfield  {author} {\bibinfo {author} {\bibfnamefont {D.~G.}\ \bibnamefont
  {Barci}}, \bibinfo {author} {\bibfnamefont {A.}~\bibnamefont {Mendoza-Coto}},
  \ and\ \bibinfo {author} {\bibfnamefont {D.~A.}\ \bibnamefont {Stariolo}},\
  }\href {\doibase 10.1103/PhysRevE.88.062140} {\bibfield  {journal} {\bibinfo
  {journal} {Phys. Rev. E}\ }\textbf {\bibinfo {volume} {88}},\ \bibinfo
  {pages} {062140} (\bibinfo {year} {2013}{\natexlab{b}})}\BibitemShut
  {NoStop}%
\bibitem [{\citenamefont {de~Gennes}\ and\ \citenamefont
  {Prost}(1998)}]{deGPr1998}%
  \BibitemOpen
  \bibfield  {author} {\bibinfo {author} {\bibfnamefont {P.~G.}\ \bibnamefont
  {de~Gennes}}\ and\ \bibinfo {author} {\bibfnamefont {J.}~\bibnamefont
  {Prost}},\ }\href@noop {} {\emph {\bibinfo {title} {The {P}hysics of {L}iquid
  {C}rystals}}}\ (\bibinfo  {publisher} {Oxford University Press},\ \bibinfo
  {year} {1998})\BibitemShut {NoStop}%
\bibitem [{\citenamefont {Chaikin}\ and\ \citenamefont
  {Lubensky}(1995)}]{ChLu1995}%
  \BibitemOpen
  \bibfield  {author} {\bibinfo {author} {\bibfnamefont {P.~M.}\ \bibnamefont
  {Chaikin}}\ and\ \bibinfo {author} {\bibfnamefont {T.~C.}\ \bibnamefont
  {Lubensky}},\ }\href@noop {} {\emph {\bibinfo {title} {Principles of
  Condensed Matter Physics}}}\ (\bibinfo  {publisher} {Cambridge University
  Press},\ \bibinfo {year} {1995})\BibitemShut {NoStop}%
\bibitem [{\citenamefont {Brazovskii}(1975)}]{Br1975}%
  \BibitemOpen
  \bibfield  {author} {\bibinfo {author} {\bibfnamefont {S.~A.}\ \bibnamefont
  {Brazovskii}},\ }\href@noop {} {\bibfield  {journal} {\bibinfo  {journal}
  {Sov. Phys. JETP}\ }\textbf {\bibinfo {volume} {41}},\ \bibinfo {pages} {85}
  (\bibinfo {year} {1975})}\BibitemShut {NoStop}%
\bibitem [{\citenamefont {Toner}\ and\ \citenamefont
  {Nelson}(1981)}]{ToNe1981}%
  \BibitemOpen
  \bibfield  {author} {\bibinfo {author} {\bibfnamefont {J.}~\bibnamefont
  {Toner}}\ and\ \bibinfo {author} {\bibfnamefont {D.~R.}\ \bibnamefont
  {Nelson}},\ }\href {\doibase 10.1103/PhysRevB.23.316} {\bibfield  {journal}
  {\bibinfo  {journal} {Phys. Rev. B}\ }\textbf {\bibinfo {volume} {23}},\
  \bibinfo {pages} {316} (\bibinfo {year} {1981})}\BibitemShut {NoStop}%
\bibitem [{\citenamefont {Barci}\ and\ \citenamefont
  {Stariolo}(2007)}]{BaSt2007}%
  \BibitemOpen
  \bibfield  {author} {\bibinfo {author} {\bibfnamefont {D.~G.}\ \bibnamefont
  {Barci}}\ and\ \bibinfo {author} {\bibfnamefont {D.~A.}\ \bibnamefont
  {Stariolo}},\ }\href {\doibase 10.1103/PhysRevLett.98.200604} {\bibfield
  {journal} {\bibinfo  {journal} {Phys. Rev. Lett.}\ }\textbf {\bibinfo
  {volume} {98}},\ \bibinfo {eid} {200604} (\bibinfo {year}
  {2007})}\BibitemShut {NoStop}%
\bibitem [{\citenamefont {Sornette}(1987)}]{So1987}%
  \BibitemOpen
  \bibfield  {author} {\bibinfo {author} {\bibfnamefont {D.}~\bibnamefont
  {Sornette}},\ }\href@noop {} {\bibfield  {journal} {\bibinfo  {journal} {J.
  Physique}\ }\textbf {\bibinfo {volume} {48}},\ \bibinfo {pages} {151}
  (\bibinfo {year} {1987})}\BibitemShut {NoStop}%
\bibitem [{\citenamefont {Kosterlitz}\ and\ \citenamefont
  {Thouless}(1973)}]{KoTh1973}%
  \BibitemOpen
  \bibfield  {author} {\bibinfo {author} {\bibfnamefont {J.~M.}\ \bibnamefont
  {Kosterlitz}}\ and\ \bibinfo {author} {\bibfnamefont {D.~J.}\ \bibnamefont
  {Thouless}},\ }\href@noop {} {\bibfield  {journal} {\bibinfo  {journal} {J.
  Phys. C: Solid State Physics}\ }\textbf {\bibinfo {volume} {6}},\ \bibinfo
  {pages} {1181} (\bibinfo {year} {1973})}\BibitemShut {NoStop}%
\bibitem [{\citenamefont {Mermin}\ and\ \citenamefont
  {Wagner}(1966)}]{MeWa1966}%
  \BibitemOpen
  \bibfield  {author} {\bibinfo {author} {\bibfnamefont {N.~D.}\ \bibnamefont
  {Mermin}}\ and\ \bibinfo {author} {\bibfnamefont {H.}~\bibnamefont
  {Wagner}},\ }\href {\doibase 10.1103/PhysRevLett.17.1133} {\bibfield
  {journal} {\bibinfo  {journal} {Phys. Rev. Lett.}\ }\textbf {\bibinfo
  {volume} {17}},\ \bibinfo {pages} {1133} (\bibinfo {year}
  {1966})}\BibitemShut {NoStop}%
\bibitem [{\citenamefont {Maier}\ and\ \citenamefont
  {Schwabl}(2004)}]{MaSc2004}%
  \BibitemOpen
  \bibfield  {author} {\bibinfo {author} {\bibfnamefont {P.~G.}\ \bibnamefont
  {Maier}}\ and\ \bibinfo {author} {\bibfnamefont {F.}~\bibnamefont
  {Schwabl}},\ }\href {\doibase 10.1103/PhysRevB.70.134430} {\bibfield
  {journal} {\bibinfo  {journal} {Phys. Rev. B}\ }\textbf {\bibinfo {volume}
  {70}},\ \bibinfo {pages} {134430} (\bibinfo {year} {2004})}\BibitemShut
  {NoStop}%
\bibitem [{\citenamefont {Barci}\ and\ \citenamefont
  {Stariolo}(2009)}]{BaSt2009}%
  \BibitemOpen
  \bibfield  {author} {\bibinfo {author} {\bibfnamefont {D.~G.}\ \bibnamefont
  {Barci}}\ and\ \bibinfo {author} {\bibfnamefont {D.~A.}\ \bibnamefont
  {Stariolo}},\ }\href {\doibase 10.1103/PhysRevB.79.075437} {\bibfield
  {journal} {\bibinfo  {journal} {Phys. Rev. B}\ }\textbf {\bibinfo {volume}
  {79}},\ \bibinfo {eid} {075437} (\bibinfo {year} {2009})}\BibitemShut
  {NoStop}%
\bibitem [{\citenamefont {Nicolao}\ and\ \citenamefont
  {Stariolo}(2007)}]{NiSt2007}%
  \BibitemOpen
  \bibfield  {author} {\bibinfo {author} {\bibfnamefont {L.}~\bibnamefont
  {Nicolao}}\ and\ \bibinfo {author} {\bibfnamefont {D.~A.}\ \bibnamefont
  {Stariolo}},\ }\href@noop {} {\bibfield  {journal} {\bibinfo  {journal}
  {Phys. Rev. B .}\ }\textbf {\bibinfo {volume} {76}},\ \bibinfo {eid} {054453}
  (\bibinfo {year} {2007})}\BibitemShut {NoStop}%
\bibitem [{\citenamefont {D\'{\i}az-M\'endez}\ \emph {et~al.}(2011)\citenamefont
  {D\'{\i}az-M\'endez}, \citenamefont {Mendoza-Coto}, \citenamefont {Mulet},
  \citenamefont {Nicolao},\ and\ \citenamefont {Stariolo}}]{DiMeMuNiSt2011}%
  \BibitemOpen
  \bibfield  {author} {\bibinfo {author} {\bibfnamefont {R.}~\bibnamefont
  {D\'{\i}az-M\'endez}}, \bibinfo {author} {\bibfnamefont {A.}~\bibnamefont
  {Mendoza-Coto}}, \bibinfo {author} {\bibfnamefont {R.}~\bibnamefont {Mulet}},
  \bibinfo {author} {\bibfnamefont {L.}~\bibnamefont {Nicolao}}, \ and\
  \bibinfo {author} {\bibfnamefont {D.~A.}~\bibnamefont {Stariolo}},\ }\href
  {\doibase 10.1140/epjb/e2011-20185-y} {\bibfield  {journal} {\bibinfo
  {journal} {Eur. Phys. J. B}\ }\textbf {\bibinfo {volume} {81}},\ \bibinfo
  {pages} {309} (\bibinfo {year} {2011})}\BibitemShut {NoStop}%
\end{thebibliography}

%

\end{document}


\begin{center}
{\large \bf Supplemental information}

\vspace{0.5cm}

{\large \bf Nature of long range order in stripe forming systems with long range repulsive interactions}

\vspace{0.5cm}

{\bf Alejandro Mendoza-Coto, Daniel A. Stariolo and Lucas Nicolao}
\end{center}

\vspace{0.8cm}

Here we show the fundamental steps which lead to the effective positional and orientational Hamiltonians, and discuss the main 
approximations considered.

Our starting point is the following coarse grained Hamiltonian for a scalar field $\phi(\vec{x})$ containing competing interactions and a local potential:
\begin{equation}
\mathcal{H}[\phi(\vec{x})]=\frac{1}{2}\int d^2 x\,\left(\vec{\nabla}\phi(\vec{x}) \right)^2
+\frac{1}{2}\int d^2x\int d^2x'\ \phi(\vec{x}){J}(\vec{x}-\vec{x'})\phi(\vec{x'})
+ \frac{1}{2\beta}\int d^2x\ V(\phi(\vec{x})),
\label{Ham}
\end{equation}
where $\beta=1/k_BT$ and $V(\phi)=-\frac{r}{2}\phi^2+\frac{u}{4}\phi^4$. This particular form used for the local potential is not essential for
the conclusions of our study, we chose this particular form of $V(\phi)$ for simplicity in the numerical simulations.

The non-local interaction considered by us corresponds to a kernel of the form $J(\vec{x})=J/\vert\vec{x}\vert^\alpha$, and as a consequence the interacting part 
of the original Hamiltonian can be written in Fourier space in the form:
\begin{equation}
 \mathcal{H}_i[\phi(\vec{x})]=\frac{1}{2}\int\frac{d^2k}{(2\pi)^2}A(k)\hat{\phi}(\vec{k})\hat{\phi}(-\vec{k}),
 \label{Hi}
\end{equation}
where $A(k)$ represents the fluctuation spectrum, which is isotropic and given by $A(k)=k^2+JC(\alpha)k^{\alpha-2}$, where $k$ stands for the modulus of the vector 
$\vec{k}$. Here we have used the well known result:
\begin{equation}
 \mathcal{F}\left[\frac{1}{r^\alpha}\right]=C(\alpha)k^{\alpha-2}
\end{equation}
where $C(\alpha)=2^{2-\alpha}\Gamma\left(\frac{2-\alpha}{2}\right)/\Gamma\left(\frac{\alpha}{2}\right)$ and $\mathcal{F}[...]$ represents
the Fourier transform operator (see e.g. Ref. \onlinecite{BaRiSt2013} where a detailed discussion is present, and references therein). For $\alpha <4$
the function $A(k)$ has a minimum at a finite value of wave vector $k_m$ \cite{BaRiSt2013}. 

It is known that in the class of systems described by the effective Hamiltonian (\ref{Ham}) the competition between the short range attractive interaction and the long range repulsive interaction 
gives rise to a modulation in the order parameter in the form of stripes, at least at zero external field. We now proceed to analyze local fluctuations
of a typical stripe pattern.

Locally, one can assume without loss of generality that the modulation profile extends in the ``$x$'' direction. Then, the scalar order parameter can be written in the general form 
$\phi(\vec{x})=\sum_{n\geq1} \phi_n \cos(nk_0x)$, where the amplitudes $\phi_n$ and $k_0$ are parameters which minimze the free energy. In particular, note that $k_0$ will be in general a temperature
dependent parameter proportional to the inverse modulation length of the stripe solution. 
We can now study the energy cost of perturbing such a solution by a fluctuation $u(\vec{x})$ in the phase of the modulation. 
The modulation profile changes to:
\begin{equation}
 \phi(\vec{x})=\sum_{n\geq1} \phi_n \cos(nk_0x+nk_0u(\vec{x})).
\end{equation}
In the small fluctuations limit the Fourier transform of $\phi$ is given by:
\begin{equation}
 \hat{\phi}(\vec{k})=\sum_{n\geq1} \phi_n \frac{(2\pi)^2}{2}\left(\delta(\vec{k}-n\vec{k}_0)+\delta(\vec{k}+n\vec{k}_0)\right)
 -\sum_{n\geq1} \phi_n\  \frac{nk_0}{2}\left(u(\vec{k}-n\vec{k}_0)-u(\vec{k}+n\vec{k}_0)\right),
\end{equation}
where $\vec{k}_0=k_0\vec{e}_x$. Substituting this form of $\phi$ back into equation (\ref{Hi}) and neglecting zero average terms we get for the
leading contribution:
\begin{equation}
 \mathcal{H}_i=\frac{V}{4}\sum_{n\geq1}A(nk_0)\, \phi_n^2+\frac{1}{2}\int\frac{d^2k}{(2\pi)^2}\ \mathcal{B}(\vec{k})\ \hat{u}(\vec{k})\hat{u}(-\vec{k}),
\label{He}
 \end{equation}
where 
\begin{equation}
\mathcal{B}(\vec{k})=\sum_{n\geq1}\frac{\phi_n^2n^2k_0^2}{4}(2\pi)^4\left(A(\vec{k}-n\vec{k}_0)+A(\vec{k}+n\vec{k}_0)\right),
\label{Bk}
\end{equation}
and $V$ is the system size. In the long wavelength limit ($k\rightarrow0$) the dominant contribution to the spectrum function in (\ref{Bk}) is around $\pm nk_0$,
where $k_0$ satisfies:
\begin{equation}
 \frac{\partial}{\partial k_0}\left(\sum_{n\geq1}\frac{1}{4}A(nk_0)\phi_n^2\right)=0.
\end{equation}
An expansion of (\ref{Bk}) in the long wavelength limit leads to:
\begin{equation}
\mathcal{B}(\vec{k})=\gamma_xk_x^2+\gamma_yk_y^4.
\label{eBk}
\end{equation}

Equations (\ref{He}) to (\ref{eBk}) imply that the local effective Hamiltonian 
\begin{equation}
 \mathcal{H}_l=\mathcal{H}_{0l}+\frac{1}{2}\int\frac{d^2k}{(2\pi)^2}\left(\gamma_xk_x^2+\gamma_yk_y^4\right)\hat{u}(\vec{k})\hat{u}(-\vec{k})
\label{Hul}
 \end{equation}
is of a more general nature than usually assumed. In particular, it is not restricted to a single mode approximation which should limit its validity only 
to a neighborhood of the critical point. This result also implies
that although inclusion of higher order modes enhances the values of the elastic coefficients 
$\gamma_x$ and $\gamma_y$ , this growing stiffness will not be able to drive the system
to a state with long range or quasi long range positional order. 

To proceed further, it is important to note that the previous expansion of $\mathcal{B}(\vec{k})$ in the long wave 
limit actually represents small deviations in 
the local wave vector of the modulation ($\vec{k}_0+k_0\vec{\nabla}u(\vec{x})$) from its optimum value. This means that regions far away from 
$\vec{k}_0$ were neglected in the original spectrum $A(k)$ in equation (\ref{Hi}). In particular, energy contributions
coming from the long range tail of the repulsive interaction were left out from the previous analysis.

\begin{figure}[ht!]
\includegraphics[scale=0.6]{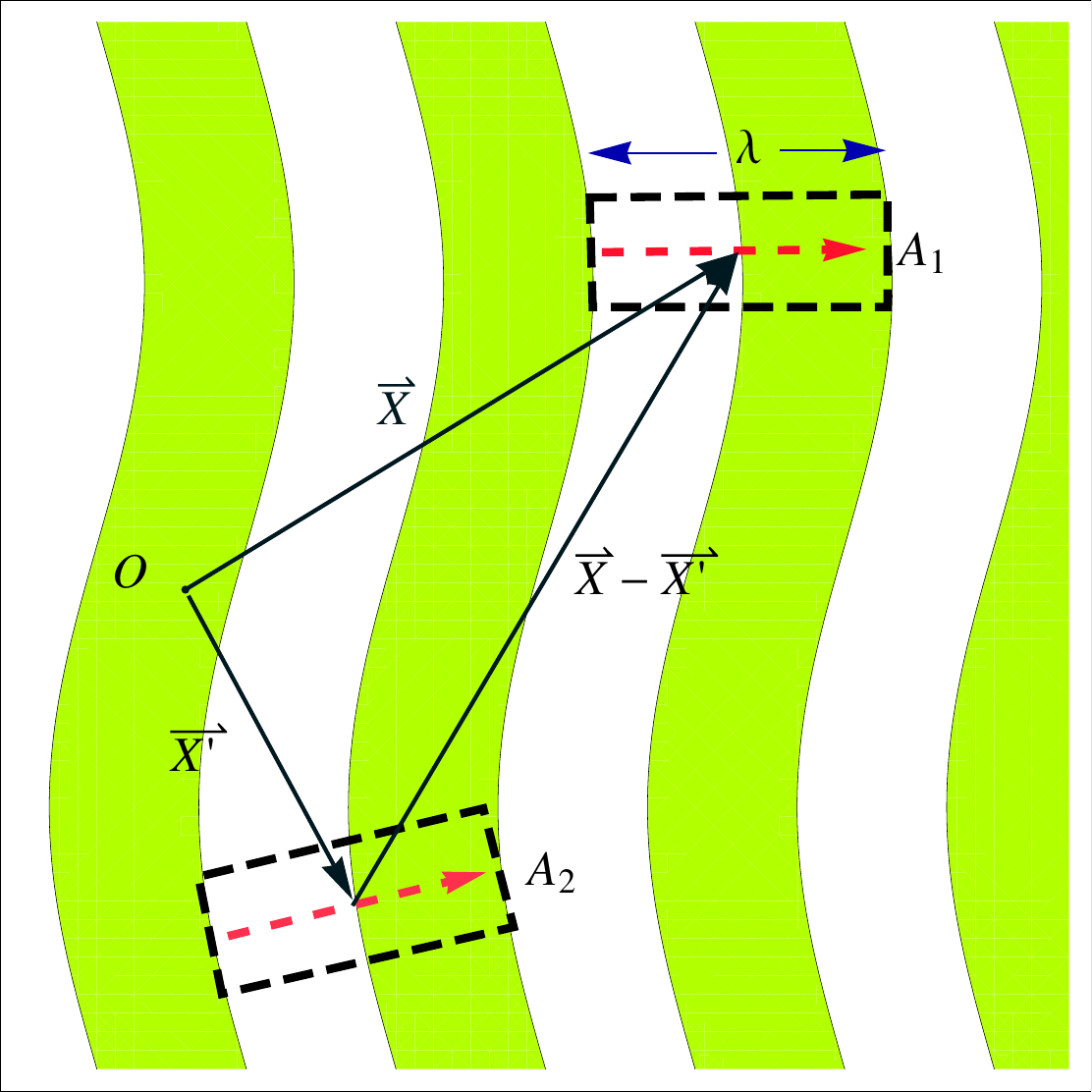}
\caption{Schematic representation of the long range interaction between two elementary stripe dipoles.}
\label{dipoles}
\end{figure}

We go on to the calculation of the explicit interaction between far apart domains (basic cells), considering the basic cell 
as a region containing a single interface with average length equal to the modulation length $\lambda$, as depicted in Figure \ref{dipoles}.

The interaction between a pair of dipolar cells is given by:
\begin{equation}
 \delta \mathcal{H}_{nl}=\frac{J}{2}\int_{\delta A_1} d^2x\int_{\delta A_2} d^2x'\ \frac{\phi(\vec{x})\phi(\vec{x'})}
 {\vert\vec{x}-\vec{x'}\vert^\alpha}.
\label{lrange}
\end{equation}
In the limit of interest, $\vert\vec{x}-\vec{x'}\vert\gg\lambda$, the integral can be calculated to any order by a multipolar expansion.
Keeping terms up to second order we find:
\begin{equation}
 \delta \mathcal{H}_{nl}=\frac{J}{2}\left(\frac{Q_1Q_2}{\vert\vec{x}_{21}\vert^{\alpha}}-
 \alpha\frac{Q_2\vec{x}_{21}\cdot\vec{P}_1}{\vert\vec{x}_{21}\vert^{\alpha+2}}+
 \alpha\frac{Q_1\vec{x}_{21}\cdot\vec{P}_2}{\vert\vec{x}_{21}\vert^{\alpha+2}}+
 \alpha\frac{\vec{P}_2\cdot\vec{P}_1}{\vert\vec{x}_{21}\vert^{\alpha+2}}-\alpha(\alpha+2)
 \frac{(\vec{P}_2\cdot\vec{x}_{21})(\vec{P}_1\cdot\vec{x}_{21})}{\vert\vec{x}_{21}\vert^{\alpha+2}}\right)\delta A_1\delta A_2,
\end{equation}
where $\vec{x}_{21}$ represents $\vec{x}-\vec{x}'$, $Q_i=(1/\delta A_i)\int_{\delta A_i}d^2x\ \phi(\vec{x})$ is a generalized ``charge'' and 
$\vec{P}_i=(1/\delta A_i)\int_{\delta A_i}d^2x\ \vec{x}\ \phi(\vec{x})$ is a generalized ``dipole moment''. Then, up to second order, the interaction
between two basic cells is composed by a term like a charge-charge interaction, two terms of type dipole-charge interaction and a generalized 
dipole-dipole type interaction. 

The total contribution of this interaction energy in the continuum limit is:
\begin{eqnarray}
\nonumber
 \mathcal{H}_{nl}&=&\frac{J}{2}\int d^2x_1d^2x_2\left(\frac{Q(\vec{x}_1)Q(\vec{x}_2)}{\vert\vec{x}_{21}\vert^{\alpha}}-
 \alpha\frac{Q(\vec{x}_2)\vec{x}_{21}\cdot\vec{P}(\vec{x}_1)}{\vert\vec{x}_{21}\vert^{\alpha+2}}+
 \alpha\frac{Q(\vec{x}_1)\vec{x}_{21}\cdot\vec{P}(\vec{x}_2)}{\vert\vec{x}_{21}\vert^{\alpha+2}}\right. \\
 &+& \left. \alpha\frac{\vec{P}(\vec{x}_2)\cdot\vec{P}(\vec{x}_1)}{\vert\vec{x}_{21}\vert^{\alpha+2}}-\alpha(\alpha+2)
 \frac{(\vec{P}(\vec{x}_2)\cdot\vec{x}_{21})(\vec{P}(\vec{x}_1)\cdot\vec{x}_{21})}{\vert\vec{x}_{21}\vert^{\alpha+2}}\right)\Omega(\vert\vec{x}_{21}\vert),
\end{eqnarray}
where $\Omega(\vert\vec{x}_{21}\vert)$ is a short distance cutoff.
Now, translational invariance can be invoked to conclude that 
$\langle Q(\vec{x}_2)\vec{P}(\vec{x}_1)\rangle=\langle Q(\vec{x}_1)\vec{P}(\vec{x}_2)\rangle$, where $\langle \cdots \rangle$ means a
thermal average. 
This implies that the ``dipole-charge'' interactions do not contribute to the free energy. 
In other words, charge and dipolar degrees of freedom are decoupled, and because of this the
``charge-charge'' interactions are irrelevant for the orientational order. 
Thus, the effective Hamiltonian with the degrees of freedom responsible for orientational order reduces to:
\begin{equation}
\mathcal{H}_{nl}=\frac{J}{2}\int d^2x\int d^2x'\alpha P(\vec{x})P(\vec{x}')\Omega(\vert\vec{x}-\vec{x}'\vert)\left(\frac{\vec{e}(\vec{x})\cdot\vec{e}(\vec{x}')}
{\vert\vec{x}-\vec{x}'\vert^{\alpha+2}} 
 - (\alpha+2)\frac{\vec{e}(\vec{x})\cdot(\vec{x}-\vec{x}')\vec{e}(\vec{x}')\cdot(\vec{x}-\vec{x}')}{\vert\vec{x}-\vec{x}'\vert^{\alpha+4}}\right),
\label{Hnl}
\end{equation}
which is equation (4) in the letter, where $\vec{e}(\vec{x})=\vec{P}(\vec{x})/P(\vec{x})$ represents the unit vector pointing in the 
direction of the dipolar moment of a given cell. 
The directors can be written in terms of the displacement fields $u(\vec{x})$ as $\vec{e}(\vec{x})\approx\frac{\vec{k}_0}{k_0}+\vec{\nabla}u(\vec{x})$ and
then, to the leading order in $u(\vec{x})$  Eq. (\ref{Hnl}) can be written as:
\begin{equation}
\nonumber
\mathcal{H}_{nl}=\Delta\mathcal{H}_{0nl}+\frac{J}{2}\alpha\int d^2x\int d^2x'P(\vec{x})P(\vec{x}')\Omega(\vert\vec{x}-\vec{x}'\vert)\left(\frac{\partial_yu(\vec{x})\partial_{y'}u(\vec{x}')}
{\vert\vec{x}-\vec{x}'\vert^{\alpha+2}} -(\alpha+2)\frac{(y-y')^2\partial_yu(\vec{x})\partial_{y'}u(\vec{x}')}{\vert\vec{x}-\vec{x}'\vert^{\alpha+4}}\right).
\end{equation}

Up to this point, the dipolar moments in the orientational effective Hamiltonian can fluctuate both in orientation and in magnitude. 
It is worth to note that in this kind of systems a disordered stripe structure or ``liquid of stripes'' develops at energies or temperatures larger
than the establishment of orientational or translational order. After phase separation has been settled in the form of a liquid of stripes,
 the stiffness of the stripes in the longitudinal direction greatly exceeds the stiffness in the transversal direction. 
Actually, this is a reflection of the existence of Goldstone soft modes in the transversal direction, evident in the isotropic nature of the
spectrum of fluctuations at wave vector $k_m$ in the stripe liquid state, which plays 
a major role in the kind of order at low temperature in these systems. The possible order-disorder transition and phase separation in these stripe systems
are in different energy scales. It is around the phase separation transition 
point that fluctuations in the width of the stripes (magnitude of generalized dipolar moments) are important.
At lower temperatures, when the liquid stripe structure has already been settled, longitudinal fluctuations are
very small. Then, in order to study the nature of orientational order,  it seems justified to consider $\vert\vec{P}(\vec{x})\vert$ as being constant. 
In this way we arrive at the final form of the orientational effective Hamiltonian (eq. (5) in the letter):
\begin{equation}
\Delta\mathcal{H}_{nl}=\frac{\gamma}{2}\int d^2x\int d^2x'\Omega(\vert\vec{x}-\vec{x}'\vert)\left(\frac{\partial_yu(\vec{x})\partial_{y'}u(\vec{x}')}
{\vert\vec{x}-\vec{x}'\vert^{\alpha+2}} -(\alpha+2)\frac{(y-y')^2\partial_yu(\vec{x})\partial_{y'}u(\vec{x}')}{\vert\vec{x}-\vec{x}'\vert^{\alpha+4}}\right)
\label{posH}
\end{equation}
where $\gamma=\alpha JP^2$. It is instructive to analyse the non-local effective Hamiltonian in Fourier space. It is known that
\begin{equation}
 \mathcal{F}\left[\frac{1}{r^\alpha}\right]=C(\alpha)\,k^{\alpha-2},
\end{equation}
where $C(\alpha)=2^{2-\alpha}\Gamma\left(\frac{2-\alpha}{2}\right)/\Gamma\left(\frac{\alpha}{2}\right)$ and $k=\sqrt{k_x^2+k_y^2}$.
Then it is straightforward to conclude that $ \mathcal{F}\left[\frac{1}{r^{\alpha+2}}\right]=C(\alpha+2)k^\alpha$,
and
\begin{eqnarray}
 \mathcal{F}\left[\frac{y^2}{r^{\alpha+4}}\right]&=&-C(\alpha+4)\frac{\partial^2{(k^{2+\alpha})}}{\partial k_y^2} \nonumber \\
 &=&-C(\alpha+4)(2+\alpha)k^\alpha-C(\alpha+4)(2+\alpha)\alpha k^{\alpha-2}k_y^2.
\end{eqnarray}
With these results the energy (\ref{posH}) can be written in Fourier space as:
\begin{eqnarray}
\nonumber
\Delta \mathcal{H}_{nl}&=&\frac{\gamma}{2}\int \frac{d^2 k}{(2\pi)^2} \left[C(\alpha+2)k^\alpha+C(\alpha+4)(2+\alpha)^2k^\alpha\right.\\
 &+&\left. C(\alpha+4)(2+\alpha)^2\alpha k^{\alpha-2}k_y^2\right]k_y^2\hat u(\vec{k})\hat u(-\vec{k}).
\end{eqnarray}
Noting that $C(\alpha+2)+C(\alpha+4)(2+\alpha)^2\equiv0$, the previous result simplifies to:
\begin{equation}
\Delta \mathcal{H}_{nl}=\frac{\gamma}{2}\int \frac{d^2 k}{(2\pi)^2} l\ k^{\alpha-2}k_y^4\hat u(\vec{k})\hat u(-\vec{k})
 \label{fin}
\end{equation}
where $l=C(\alpha+4)(2+\alpha)^2\alpha$. 
Finally, summing the local contribution of eq. (\ref{Hul}) with the non local one of eq. (\ref{fin}) we arrive at the
effective Hamiltonian for the positional degrees of freedom (eq. (6) of the letter):
\begin{equation}
\Delta\mathcal{H}=
\frac{1}{2}\int\frac{d^2k}{(2\pi)^2}(\gamma_xk_x^2+\gamma_yk_y^4+\gamma_{nl}k^{\alpha-2}{k_y}^4)\hat{u}(\vec{k})\hat{u}(-\vec{k})
\label{Heff}
\end{equation}
where $\gamma_{nl}=\gamma\alpha(2+\alpha)^2C(\alpha+4)$. 

This is the relevant effective Hamiltonian which determines the nature of positional order in the stripe system. The next step
is to obtain an effective Hamiltonian for the orientational degrees of freedom. Follwing the previous method of analysis, 
one can argue that the effective orientational Hamiltonian will be composed by two contributions.
The local one, that have been obtained in previous works, can be seen as the result of a coarse graining process over lengths of the order of the 
positional correlation length $\xi_u$. Beyond this lengthscale the director field of the stripes will show deviations 
which can be modelled, in the small deviation regime, by an effective orientational local Hamiltonian of the form:
\begin{equation}
\Delta\mathcal{H}_{ol}=\frac{\gamma_{ol}}{2}\int d^2x\, (\vec{\nabla}\theta)^2(\vec{x}) 
\label{Hol}
\end{equation}
where $\theta(\vec{x})$ is the angle between two neighboring domains pointing along directions $\vec n$ and $\vec n'$. Note that these
stripe domains have linear size of order $\xi_u$, much bigger than the size of elementary dipoles considered previously for the study of
positional order. The elastic coefficient 
$\gamma_{ol}$ can be estimated to be $J\alpha^2 P^2/(4\xi_u^{\alpha-2})$. On the other hand, the interaction between domains separated by distances much 
larger than $\xi_u$ can be taken into account following formally the same steps as done for the non local interaction between positional  
degrees of freedom. The effective interaction between domains will take the same form as (\ref{Hnl}):
\begin{equation}
\Delta \mathcal{H}_{onl}=\frac{\gamma}{2}\int d^2x\int d^2x'\,\Omega(\vert\vec{x}-\vec{x}'\vert)\left(\frac{\vec{n}(\vec{x})\cdot\vec{n}(\vec{x}')}
{\vert\vec{x}-\vec{x}'\vert^{\alpha+2}} 
 - (\alpha+2)\frac{\vec{n}(\vec{x})\cdot(\vec{x}-\vec{x}')\vec{n}(\vec{x}')\cdot(\vec{x}-\vec{x}')}{\vert\vec{x}-\vec{x}'\vert^{\alpha+4}}\right).
\label{Honl}
\end{equation}
where $\gamma=\alpha JP^2$. Thus, $\Delta \mathcal{H}_{o}=\Delta \mathcal{H}_{ol}+\Delta \mathcal{H}_{onl}$ is the complete orientational effective Hamiltonian. To study
the stability of an ordered orientational phase against small deviations of the local directors $\vec{n}(\vec{x})$ a spin wave analysis can be done.
In the spin wave limit the orientational Hamiltonian reduces to:
\begin{equation}
\Delta \mathcal{H}_{o}=\frac{\gamma_{ol}}{2}\int d^2x\, (\vec{\nabla}\theta)^2(\vec{x})+\frac{\gamma}{2}\int d^2x\int d^2x'\,
\Omega(\vert\vec{x}-\vec{x}'\vert)\left(\frac{\theta(\vec{x})\theta(\vec{x}')}{\vert\vec{x}-\vec{x}'\vert^{2+\alpha}}-
 (2+\alpha)\frac{(\Delta y)^2\theta(\vec{x})\theta(\vec{x}')}{\vert\vec{x}-\vec{x}'\vert^{4+\alpha}}\right).
\label{Ho}
\end{equation}
From Eq. (\ref{fin}) the Fourier transform is easily found to be:
\begin{equation}
\mathcal{H}_{o}=\frac{1}{2}\int \frac{d^2k}{(2\pi)^2}\,(\gamma_{ol}k^2+\gamma_{nl}k^{\alpha-2}{k_y}^2)\,\theta(\vec{k})\theta(-\vec{k})
\end{equation}
with $\gamma_{nl}=\gamma\alpha(2+\alpha)^2C(\alpha+4)$.

\bibliographystyle{apsrev4-1}
\bibliography{ultrathin,nematics}